Title: Kolmogorov turbulence in a random-force-driven Burgers equation

\documentstyle[12pt]{article}
\newfont{\feff}{cmti10}
\begin{document}

\input{psfig.sty}

\title{\bf Kolmogorov Turbulence in a Random Force Driven Burgers Equation}

\author{\\{Alexei Chekhlov and Victor Yakhot} \\ \\{\em Program in Applied and
Computational Mathematics,} \\ {\em Princeton University,}\\
{\em Princeton, New Jersey 08544, USA}}

\maketitle

\psfull

\begin{abstract}
The dynamics of velocity fluctuations, governed by the one-dimensional
Burgers equation, driven by a white-in-time random force $f$ with the
spacial spectrum $\overline{\left|f(k)\right|^2}\propto k^{-1}$, is
considered.  High-resolution numerical experiments conducted in this
work give the energy spectrum $E(k)\propto k^{-\beta}$ with
$\beta=\frac{5}{3}\pm 0.02$.  The observed two-point correlation
function $C(k,\omega)$ reveals $\omega \propto k^{z}$ with the
``dynamic exponent'' $z\approx 2/3$.  High-order moments of velocity
differences show strong intermittency and are dominated by powerfull
large-scale shocks. The results are compared with predictions of the
one-loop renormalized perturbation expansion.
\end{abstract}

PACS number(s): 47.27.Gs.

\vspace{0.5in}
{}From a theoretical viewpoint, one of the most challenging features of
strong hydrodynamic turbulence is an interplay between random almost
gaussian background and coherent ordered structures responsible for
deviations from gaussian statistics. Although coherent structures were
visualized in three-dimensional flows as sheets or tubes of high
vorticity \cite{3D}, little is known about their analytic structure,
stability and, as a consequence, about their relevance to turbulence
dynamics. In two-dimensional systems, the role of coherent structures
is much better understood: the flow can be decomposed into two
components, the background field having close to gaussian statistics
and coherent, extremely stable point vortices, responsible for
strongly non-gaussian features of the flow \cite{2D}. Still, the
analytic structure of the vortices and the distributions of their
sizes and strengths are not yet understood and this is why the full
statistical theory of two-dimensional turbulence does not exist.

The analytic properties of the one-dimensional Burgers equation \cite{Burgers}
\begin{eqnarray}
\frac{\partial v}{\partial t} +\frac{1}{2}\frac{\partial v^{2}}
{\partial x}=\nu_{0}\frac{\partial^{2} v}{\partial x^{2}}
\end{eqnarray}

\noindent
subject to initial and boundary conditions, are understood rather
well: the flow is dominated by the shocks, leading to $E(k)\propto
k^{-2}$ \cite{Burgers}. Moreover, in some cases, the Burgers equation
has a stationary solution. For example, if $v=-U$ and $v=U$ at
$x=\infty$ and $-\infty$ respectively, $U(x)=-U\,\tanh
\left(x\,U/(2\,\nu_0)\right)$, which describes a single shock of the
width $l\approx \nu/U$.  In this particular solution, ``fluid''
particles, created at the boundaries, are carried towards the center
of the shock where they disappear.  The shock formation is the most
significant dynamic property of the Burgers equation and the shocks
were studied in the systems decaying from some initial conditions and
driven by the large-scale random noise \cite{Sinai}. In the latter
case, the energy spectrum is $E(k)\propto k^{-2}$ and all velocity
structure functions $S_{2n}(r)=\overline{(u(x+r)-u(x))^{2n}}$ scale as
$S_{2n}(r)\propto r^1$, characteristic of the shocks.

A totally different result is found in a system governed by $(1)$
driven by a white-in-time random force $f(x,t)$ defined by its correlation
function:
\begin{eqnarray}
\overline{f(k,\omega)f(k',\omega')}=2\,(2\pi)^{2}\,D_0\,k^{-y}
\delta(k+k')\delta(\omega+\omega')
\end{eqnarray}

\noindent
with $y=-2$ corresponding to   thermal equilibrium.
Here, in the limit $k\rightarrow 0$ and $\omega\rightarrow 0$
the two-point velocity correlation function is given by
\begin{eqnarray}
C(k,\omega)=\int\int\frac{\overline{v(k,\omega)\,v(k',\omega')}}{2}
\,\frac{d\omega'}{2\pi}\,\frac{dk'}{2\pi}\propto k^{-\alpha}
C(\frac{\omega}{k^{z}})
\end{eqnarray}

\noindent
with $\alpha=z=3/2$ corresponding to $E(k)$=const.  Both exponents $z$
and $\alpha$ were evaluated from the theories based on the one-loop
renormalized perturbation expansions \cite{FNS,YS} and were confirmed
by numerical experiments \cite{YS}. In this case, the small-scale
forcing was strong enough to prevent formation of the shocks and the
$k^{-2}$-energy spectrum. In recent articles \cite{twoloop}, it was
shown that computation of the second loops for the $y=-2$ case does
not invalidate the results of the one-loop approximation.

In this work, we are interested in an intermediate case of a system
governed by (1) with the forcing function added to the right side
defined by the relation (2) with $y=1$. This case is extremely
interesting because it corresponds to the ``almost'' constant energy
flux $\Pi(k)$ in the wave-number space: $\Pi(k)\propto \log(k/k_0)$,
where $k_0$ is the inverse largest allowed scale in the system. Since
the analytic structure of $(1)$ resembles that of the Navier-Stokes
equations, the Kolmogorov argument leading to $E(k)\propto
k^{-\frac{5}{3}}$ can be applied at least on a superficial level.
However, in this case the process of the Kolmogorov spectrum
generation should compete with the natural tendency of the Burgers
equation towards formation of coherent shocks, thus leading to the
most interesting dynamics.

We investigate the fluctuations generated by the equation $(1)$ with
the more general dissipation term
$\nu_0\,(-1)^{p+1}\,\partial^{2p}v/\partial x^{2p}$ and driven by the
random force $f(x,t)$.  Numerical results, shown below, correspond to
$p=6$, which has been chosen empirically to produce sufficiently sharp
UV energy spectrum fall-off. The effect of the hyper-dissipation on
the solutions of the Burgers equation has been studied in a recent
paper \cite{Boyd} and we will not dwell upon this issue here. We will
just mention that its use is dictated by desire to have as wide a
universal range as possible and is based on the assumption that
universal IR properties do not depend on the type of dissipation
chosen. To simulate $(2)$, the random force has been assigned in the
Fourier space as: $f(k,t)=A_f/\sqrt{\delta
t}\,\left|k\right|^{-y/2}\,\sigma_k$, where $\sigma_k$ is a Gaussian
random function with $\overline{\left|\sigma_k\right|^2}=1$ and
$\delta t$ is a time-step. The force cut-off $k_{c}$ was chosen well
inside the dissipation range of the energy spectrum. In the case $p=1$
the dissipation scale is, according to the relation given above,
$l_d\approx \frac{\nu_0}{U_{0}}$ where $U_{0}$ is the rms velocity of
the most energetic shock in the system.  Spatial discretization was
based on the Fourier-Galerkin pseudospectral method with the nonlinear
term computation in the conservative form and dealiasing procedure
based on the $1/3$-rule.  Temporal discretization included two
second-order schemes: Runge-Kutta for restarting and stiffly-stable
Adams-type scheme described in \cite{Israely} for serial
computations. Spectral resolution employed here was $12288$ including
the dealiased modes. Other parameters were chosen to be:
$\nu_0=9.0\times 10^{-40}$, $\delta t=5.0\times 10^{-5}$, and
$A_f=1.4142\times 10^{-3}$. It was carefully tested that this
parameter set does lead to the strong coupling in the inertial range
$10\le k\le 600$, such that the viscous term in the energy equation
derived from $(1)$ is negligibly small compared with the corresponding
nonlinear term.

The results of numerical experiments are presented in Figs. $1-5$.
Integration was performed for over approximately $11\,\tau_{to}$,
where: $\tau_{to}=\pi/V_{rms}\approx 100$ is a largest eddy turnover
time.  After approximately $0.5\,\tau_{to}$ the statistically
steady-state has been achieved. Fig.$1$ presents $2$ successive
realizations of the velocity field in this steady-state. One can see
the typical saw-tooth structures, characteristic of the dynamical
system governed by Burgers equation. In our case, however, they are
superimposed by the random component of the velocity field.  It was
noticed that the system spends most of the time in the state where
there are few (3-4) large-amplitude shocks and many small-amplitude
ones. But processes leading to the creation of a single strong shock
and its later breakdown into several smaller ones constantly take
place. The time-evolution of the total energy in the system $E(t)$
demonstrates strong (with the amplitude of more than $100\%$ of the
average energy) fluctuations, characteristic of the instability of the
large-scale structures. The time-averaged energy spectrum $E(k,t)$
($E(t)=\int E(k,t)dk$), presented in Fig.$2$, is well approximated by
the Kolmogorov law: $E(k)\propto k^{-\beta}$ with $\beta=\frac{5}{3}
\pm 0.02$. The error bars were estimated in the following way: various
values of parameter $\beta$ were used to plot the compensated energy
spectrum $e(k)=k^{\beta}E(k)$ and only the values of exponent $\beta$,
for which $e(k)$ was inside the experimental noise in the entire
interval $10<k<600$, were chosen as representable. The velocity
structure functions $S_{2n}(r)=\overline{(v(x+r)-v(x))^{2n}}$ are
shown in Fig.$3$ for $n=2-4$. We can see that all $S_{2n}(r)\propto
r^{\beta_{2n}}$ with $\beta_{2n}\approx 0.91$ indicating that these
correlation functions are dominated by coherent shocks. We were not
able to detect the logarithmic corrections to the energy spectrum
$E(k)$. However, the fact that the high-order moments, presented on
Fig.3 are characterized by the exponents close but not exactly equal
to unity, indicate that the logarithmic contributions cannot be ruled
out. One remarkable result is related to the dissipation rate
correlation function presented in Fig.$4$,
G(r)=$\overline{\epsilon(x+r)\epsilon(x)}\propto r^{-\mu}$, with
intermittency exponent $\mu\approx 0.2\pm 0.05$ measured inside the
universal range $0.01\le r\le 0.63$.  Note that the dissipation rate
correlation function is defined in the physical space and the
$1/3$-rule dealiasing procedure was also used for its computation.
The obtained value of the dissipation rate exponent $\mu$ is close to
that observed in the experiments on 3-dimensional turbulence:
$\mu=0.25\pm0.05$, see \cite{experiment}, and its general shape
resembles the model shape of $G(r)$ proposed for 3-dimensional
turbulence in \cite{Nelkin}. We would like to emphasize that in the
present work the dissipation rate
$\epsilon(x)=\nu_{0}(\frac{\partial^{p} v}{\partial x^{p}})^{2}$ with
$p=6$ strongly differs from the normal viscosity case with $p=1$. The
fact that the exponent $\mu$ obtained in this work was close to the
one observed in real-life turbulence, provides an indication that the
correlation function of the dissipation rate for the inertial range
values of the displacement $r$ is independent on the structure of the
dissipation range. A similar conclusion was reached from the recent
numerical experiments of three-dimensional turbulence in
\cite{Vadim}. As one can see from Fig.4, the accuracy of the exponent
$\mu$ is not as good as of the exponent in the expression for the
energy spectrum.  In addition, to assess importance of the result, the
role of the hyperviscosity in the dissipation rate correlation
function is to be further investigated.

Important information about the dynamics of a non-linear system can be
extracted from the correlation function defined by (3). In the
scale-invariant regime, according to the theories based on the
one-loop renormalized perturbation expansion \cite{FNS,YS}
\begin{eqnarray}
C(k,\omega)=\frac{D_0\,k^{-1}}{\omega^2 + \nu^2(k,\omega=0)\,k^4},
\end{eqnarray}

\noindent
where $\nu(k,\omega=0)\approx
(\frac{3D_{0}}{4\pi})^{\frac{1}{3}}k^{-\frac{4}{3}}$, corresponding to
$z=2/3$ and $\alpha=-7/3$ in (3). The frequency dependence of the
effective viscosity $\nu(k,\omega)$ is neglected in the relation
$(4)$.  This is the direct consequence of the assumption that the
dynamics of the inertial range modes $v(k,\omega)$ are dominated by
the ``distant interactions'' with the modes $v(q,\Omega)$ with
$\left|k\right|\ll \left|q\right|$ and can be decribed by the
$k$-dependent eddy-viscosity.  It is clear that this approximation can
not be valid when we are interested in behaviour of the most powerfull
large-scale structures, because of the strong interaction between them
leading to the events of the shock instability observed in this
work. The energy spectrum derived from $(4)$ is: $E(k)=\int
C(k,\omega) d~\omega/(2\pi)=\left(\pi\,D_0^2/6\right)^
{1/3}\,k^{-5/3}$. The energy flux in the wave-number space can be
expressed in terms of the amplitude of the force correlation function
$D_0$ as folows: $\Pi(k)=\Pi(k_0)+2D_{0}\,log(k/k_0)$. Then, the value
of the ``Kolmogorov'' constant arising from $(6)$ is:
$C_K=\left((\Pi(k)-\Pi(k_0))/log\left(k/k_0\right)
\right)^{-2/3}\,k^{5/3}\,E(k)=\left(\pi/6\right)^{1/3}$.
The numerical value $C_{K}\approx 0.806$ agrees well with the results
of numerical simulation, see Fig.$2$. As in many other cases,
understanding of the reasons for a good agreement between the
theory, based on the one-loop renormalized perturbation expansion, and
experimental data, remains a major challenge.

The computational procedure for $C(k,\omega)$ included the following
steps. Starting from some initial moment $t=t_0$ well inside the
statistically steady-state, the solution $v(k,t)$ was stored at the
time instants $t_j=t_0+T\,j/M$, where $T=100$ was chosen to be of the
order of $\tau_{to}$.  Then at $t=t_M$ $v(k,\omega)$ was found via
discrete Fourier transform. Repeating such procedure with time and
assuming each realization of $v(k,\omega)$ to be independent of
others, which certainly is only an approximation, one can compute
$C(k,\omega)$ as an average over such realizations.  Memory
requirements allowed us to keep only $200$ first wavevectors and limit
ourselves with $M=3000$. Results of computations of $C(k,\omega)$,
presented in Fig.$5$, can be compared with the prediction $(4)$. The
relation $(4)$ was derived neglecting the infra-red divergences
resulting in the transport of the small-scale fluctuations by the
large-scale coherent structures.  This kinematic interaction
(``sweeping effect'') can be accounted for by the Doppler shift
$\omega\rightarrow \omega +kV$ in $(4)$, where $V$ is the
characteristic velocity of the large scale structures. Major interest
is whether $C(k,\omega)$ is described by $(4)$ or not and whether $V$
is zero or not. If the sweeping effect is present in the long-time
behavior then there are $3$ possible scaling regimes of $C(k,\omega)$
as $\omega\rightarrow 0+$: $C(k,\omega)\propto k^{-1}$ if $k\ll
(\omega^3/D_0)^{1/2}$, $C(k,\omega)\propto k^{-7/3}$ if
$(\omega^3/D_0)^{1/2}\ll k\ll D_0/V^3$, and $C(k,\omega)\propto
k^{-3}$ if $k\gg D_0/V^3$.  It is clear from Fig.$5$ that the
theoretical prediction $(4)$ is surprisingly accurate in both limits
of large and small frequences $\omega$. Only in a narrow intermediate
range of the wavenumbers $\omega\approx \nu(k,\omega=0)k^{2}$
prediction $(4)$ fails. The flattening of $C(k,\omega)$ observed in
this interval indicates that the scaling function $F(x)$ in $(3)$ is a
decreasing function of $x$ when $x\approx 1$.  The quantitative
agreement between theory and simulations in the limit of large
wave-numbers $k$ shows that the ``sweeping velocity'' $V$ is small.
This may be a consequence of the fact that the large-scale shocks are
almost steady.  We would also like to note that the accuracy of the
$C(k,\omega)$ computation may not be easily increased because of the
computer resources limits which have been reached by us.  The above
result leads to an interesting possibility: The infra-red divergences,
present in the theory, are not summed up into a mere transfer of
small-scale fluctuations by the large-scale structures but are
reflected in the creation of a large-scale condensate state, which in
this case has a very simple physical meaning as a collection of strong
shocks moving with a very small velocity $V$. Derivation of the
equation of motion describing dynamics of coherent shocks is an
important and interesting problem and will be the subject of future
communication.

To conclude, the above results, obtained in a simple one-dimensional
system are surprisingly similar to the outcome of experimental
investigations of the real-life three-dimensional turbulence. In the
one-dimensional case, however, the dynamics and geometrical features
of the ``turbulence'' building blocks are well understood and a
cascade process is readily envisioned as a coagulation of the weak and
wide shocks (the shock width and its amplitude are related as
$l\approx \frac{\nu_{0}}{u}$) into ever stronger and narrower
structures until the dissipation takes over. Moreover, the total
dissipation rate in an interval of the length $r$ is prescribed and is
equal to $\epsilon_r\propto ln~\frac{r\,U_{0}}{\nu_0}$. Given these
simplifications, one may hope that the full theory of Kolmogorov
turbulence in the one-dimensional Burgers equation is not out of
reach.

This work was supported by grants from ARPA and AFOSR. Many
stimulating discussions with V. Borue, R. H. Kraichnan, S. Orszag,
A. Polyakov and Ya. Sinai are gratefully aknowledged.


\begin{figure}[h]
\centerline{\psfig{file=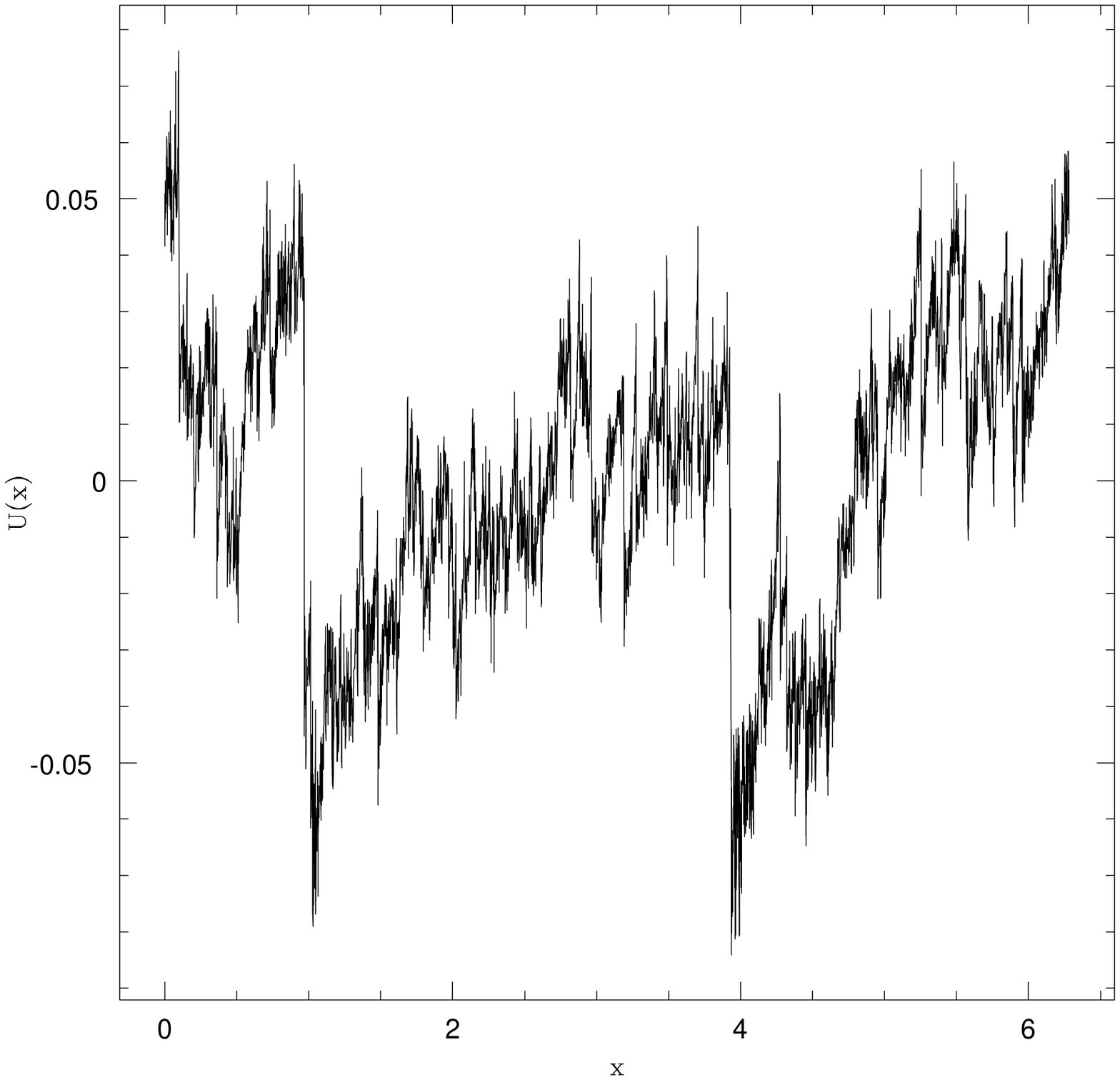,width=4.0in,height=4.0in}}
\centerline{\psfig{file=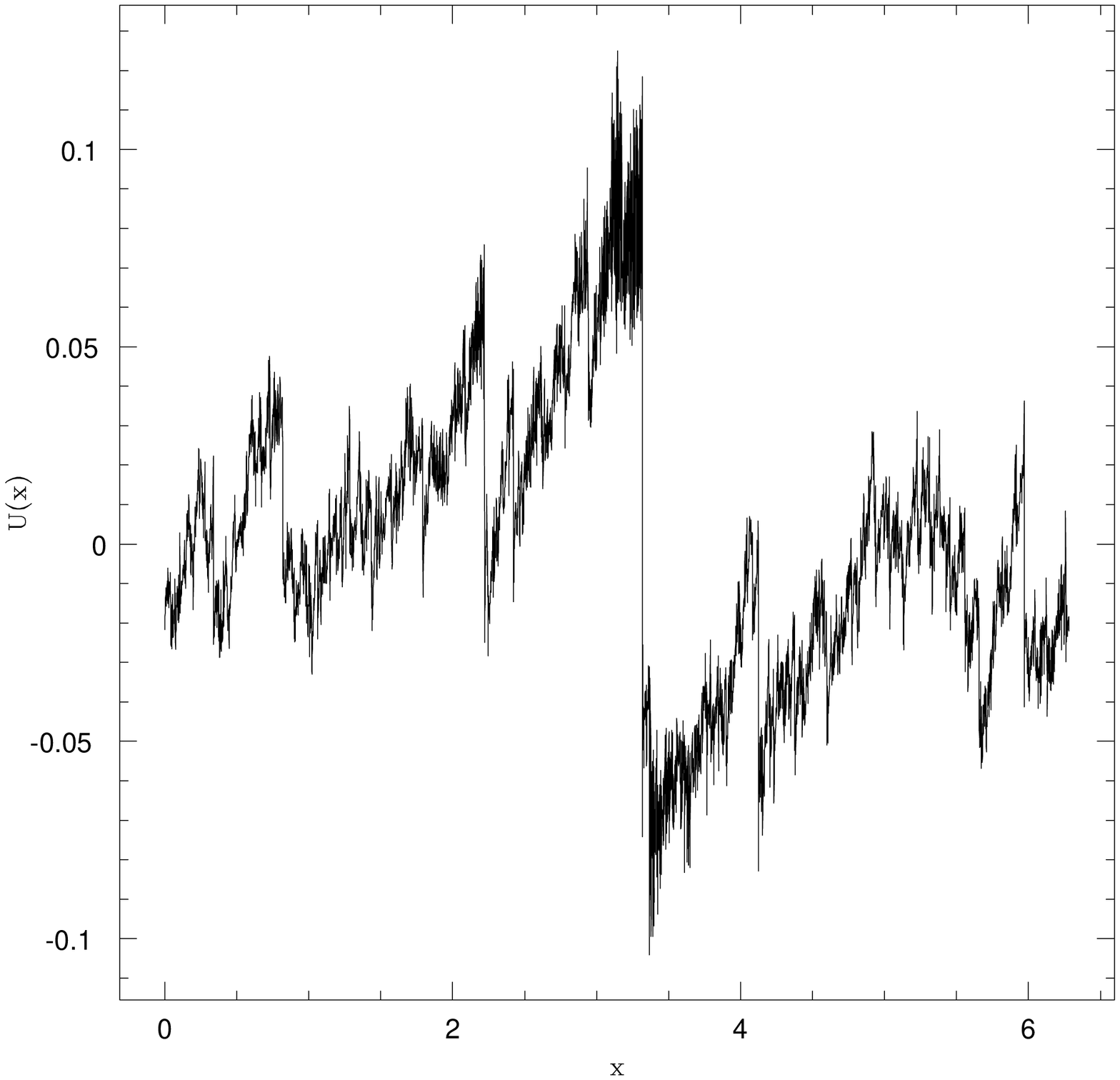,width=4.0in,height=4.0in}}
\caption{Solutions $v(x,t)$ at times $t=90$ (upper) and $t=213.5$ (lower). }
\end{figure}


\begin{figure}[h]
\centerline{\psfig{file=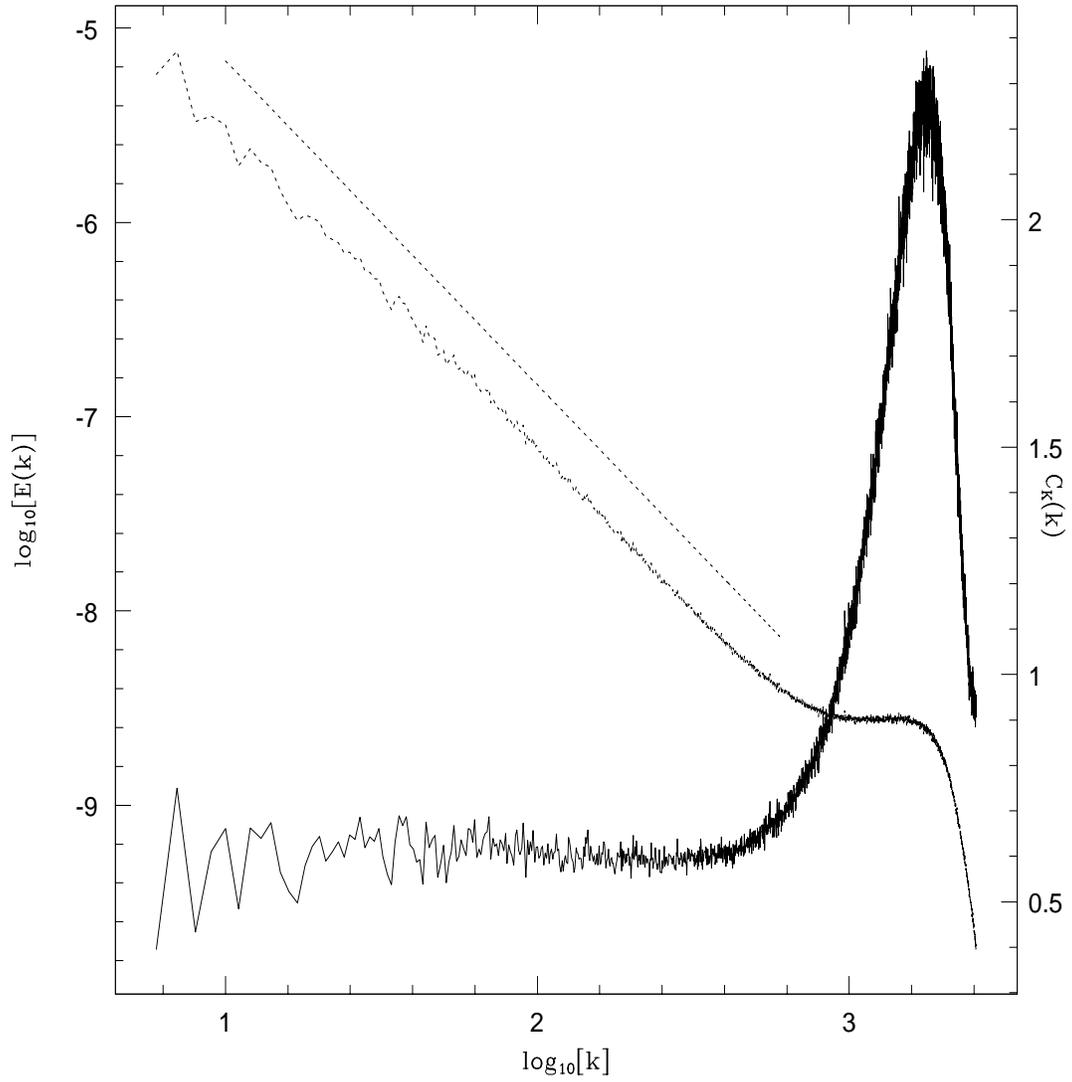,width=6.0in,height=6.0in}}
\caption{Dotted curve is the energy spectrum
$E(k)$ (left axis), the straight line above it has the exact slope $-5/3$.
Solid curve is the compensated energy spectrum $C_K$ defined in the text (right
axis).}
\end{figure}


\begin{figure}[h]
\centerline{\psfig{file=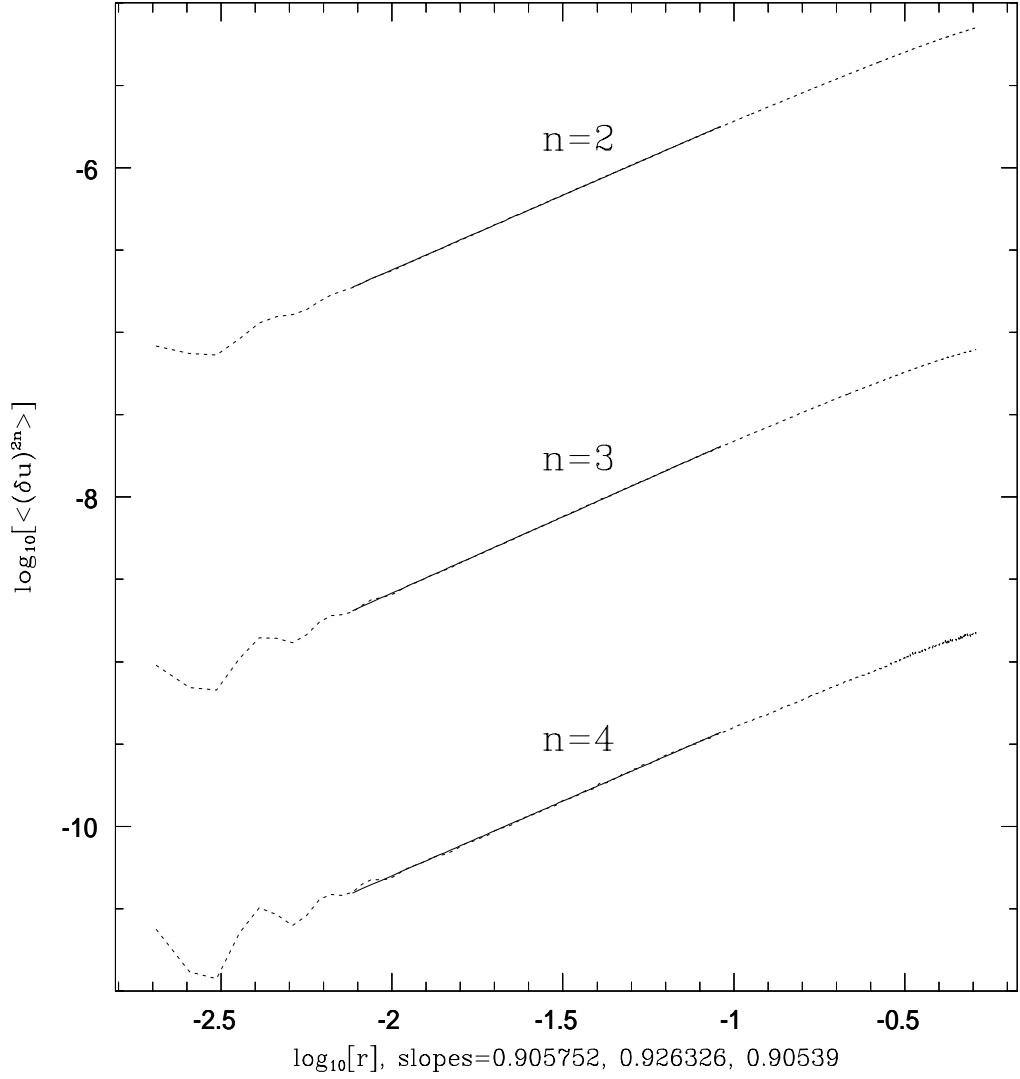,width=6.0in,height=6.0in}}
\caption{Velocity differences $\overline{\left(v(x+r)-v(x)\right)^{2n}}$ for
$n=2,3,4$ (dotted curve) with the linear least-squares fit (solid line). }
\end{figure}


\begin{figure}[h]
\centerline{\psfig{file=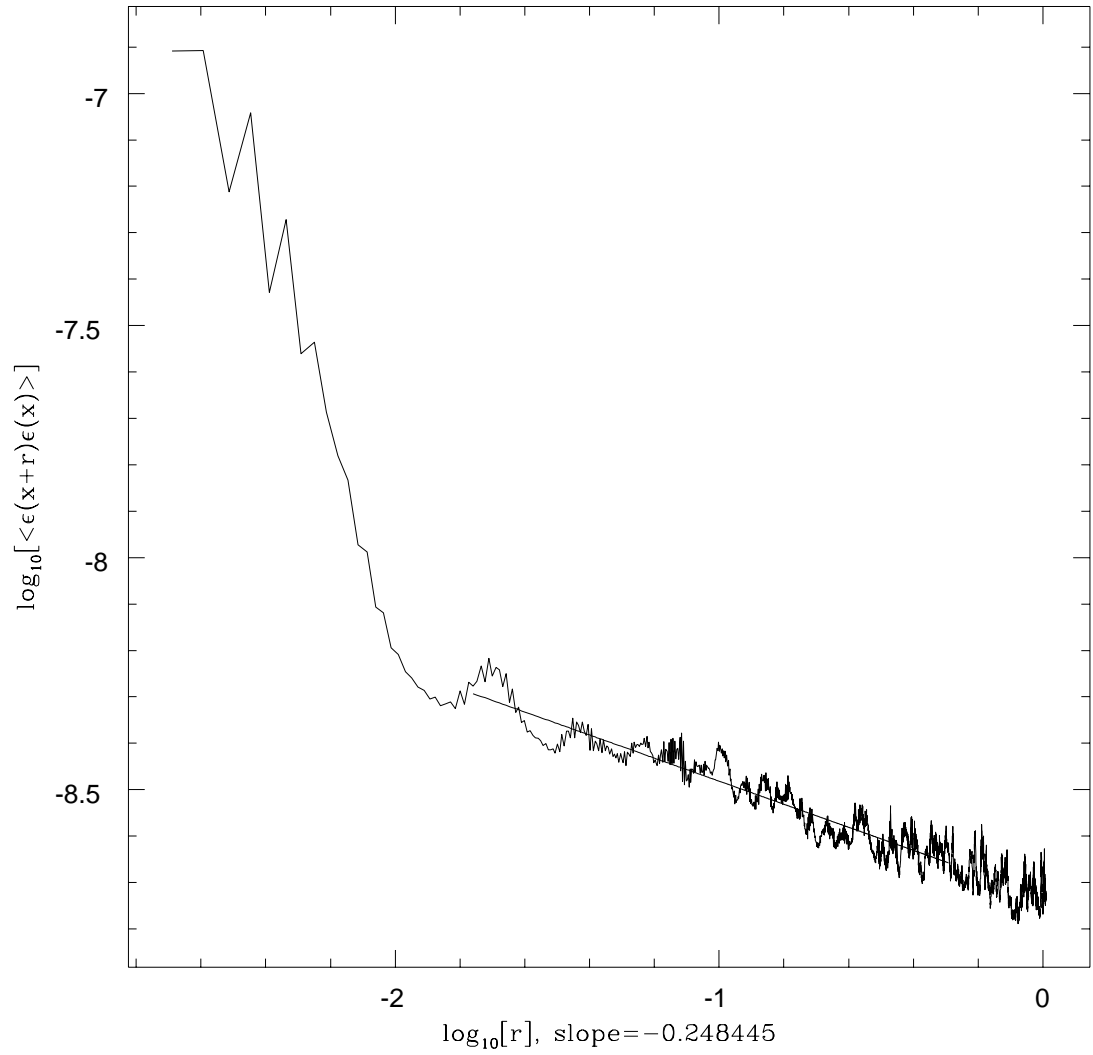,width=6.0in,height=6.0in}}
\caption{Energy dissipation correlation function
$\overline{\epsilon(x+r)\,\epsilon(x)}$ with the linear least-squares
fit, giving the intermittency exponent $\mu=0.2\pm 0.05$. }
\end{figure}


\begin{figure}[h]
\centerline{\psfig{file=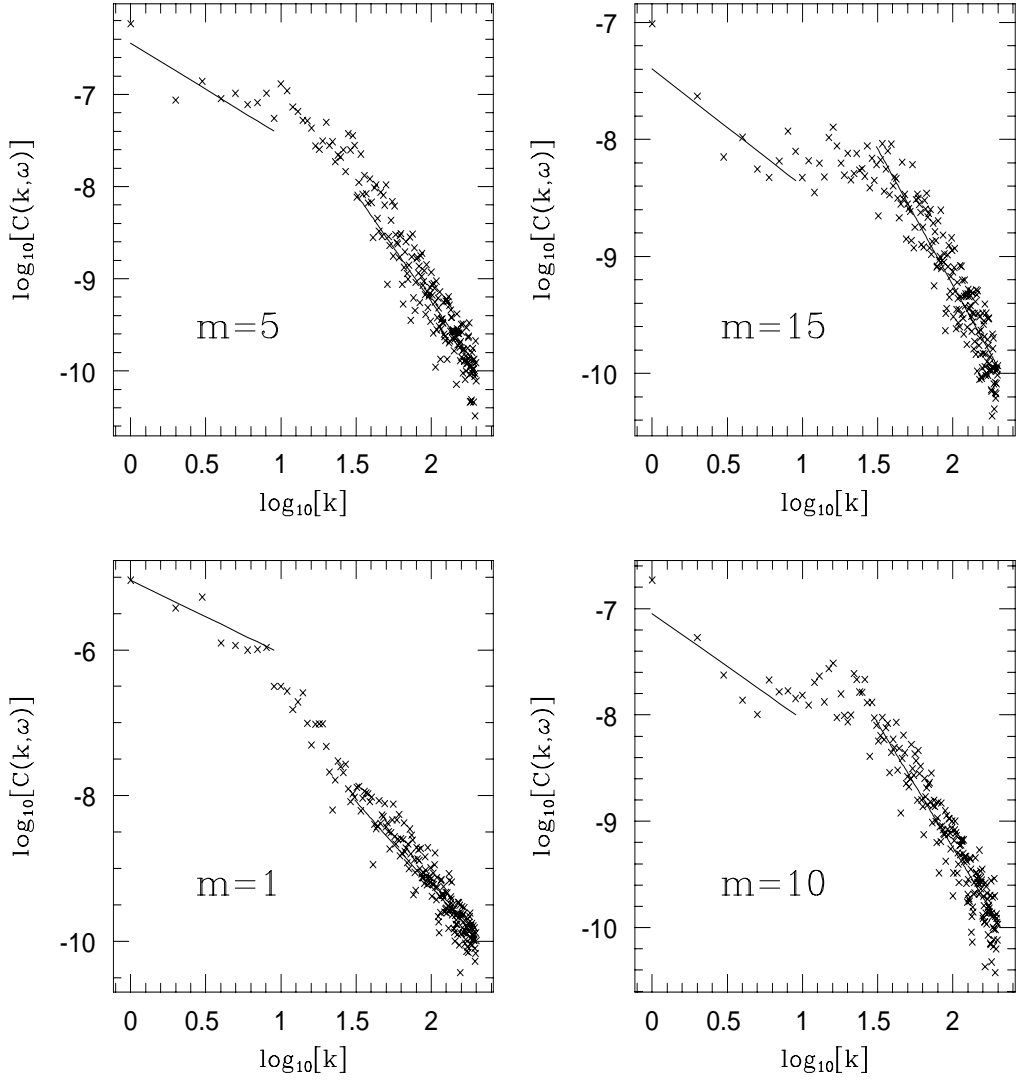,width=6.0in,height=6.0in}}
\caption{Points denote the self-correlation function of the solution
$C(k,\omega)$
for fixed time-frequencies $\omega=2\,\pi\,m/\tau$ with $m=1$,
$m=5$, $m=10$, $m=15$ and $\tau=100.05$. Solid lines denote the
corresponding asymptotics of the one-loop prediction given by $(4)$. }
\end{figure}

\end{document}